\documentclass[a4paper,11pt]{article}
\usepackage{pos}

\usepackage[sort&compress]{natbib}

\newcommand{\be}{\begin{equation}}  
\newcommand{\ee}{\end{equation}}  
\newcommand{\beq}{\begin{eqnarray}} 
\newcommand{\eeq}{\end{eqnarray}}

\newcommand{\bea}{\begin{eqnarray}}
\newcommand{\eea}{\end{eqnarray}}

\let\oldbibliography\thebibliography
\renewcommand{\thebibliography}[1]{\oldbibliography{#1}
\setlength{\baselineskip}{9.5pt}
\setlength{\itemsep}{5pt}} 

\title{Accessing proton GPDs in
asymmetric frames:
Numerical implementation}

\author*[a]{Martha Constantinou}
\emailAdd{marthac@temple.edu}
\affiliation[a]{Department of Physics,  Temple University,  Philadelphia,  PA 19122 - 1801,  USA}
\author[b]{Shohini Bhattacharya}
\affiliation[b]{Physics Department, Brookhaven National Laboratory, Upton, New York 11973, USA}
\author[c]{Krzysztof Cichy}
\affiliation[c]{Faculty of Physics, Adam Mickiewicz University, ul.\ Uniwersytetu Pozna\'nskiego 2, 61-614 Pozna\'{n}, Poland}
\author[a]{Jack Dodson}
\author[d]{Xiang Gao}
\affiliation[d]{Physics Division, Argonne National Laboratory, Lemont, IL 60439, USA}
\author[a]{Andreas Metz}
\author[b]{Swagato Mukherjee}
\author[a]{Aurora Scapellato}
\author[e]{Fernanda Steffens}
\affiliation[e]{Institut f\"ur Strahlen- und Kernphysik, Rheinische Friedrich-Wilhelms-Universit\"at Bonn,\\ Nussallee 14-16, 53115 Bonn}
\author[d]{Yong Zhao}

\abstract{In this work, we present a numerical investigation of a novel Lorentz-covariant parametrization to extract $x$-dependent GPDs using off-forward matrix elements of momentum-boosted hadrons coupled to non-local operators. The novelty of the method is the implementation of an asymmetric frame for the momentum transfer between the initial and final hadron state and the parametrization of the matrix elements into Lorentz-invariant amplitudes. The amplitudes can then be related to the standard light-cone GPDs. 

GPDs are defined in the symmetric frame, which requires a separate calculation for each value of the momentum transfer, increasing the computational cost significantly. The proposed method is powerful, as one can extract the GPDs at multiple values of the momentum transfer at the computational cost of a single value. For this proof-of-concept calculation, we use one ensemble of $N_f=2+1+1$ twisted mass fermions and a clover improvement with a pion mass of 260 MeV to calculate the proton unpolarized GPDs.}

\FullConference{%
The 39th International Symposium on Lattice Field Theory,\\
8th-13th August, 2022,\\
Rheinische Friedrich-Wilhelms-Universität Bonn, Bonn, Germany
}

\begin{document}
\maketitle

\section{Theoretical and lattice setup}
\label{sec:Lat_setup_A}

Accessing generalized parton distributions (GPDs)~\cite{Mueller:1998fv, Ji:1996ek, Radyushkin:1996nd} on the lattice is one of the crucial efforts towards understanding 3D hadron structure.
While the direct calculation of any parton distributions is impossible in Euclidean spacetime, one can construct appropriate lattice observables that can be matched to their light-cone counterparts.
Since the groundbreaking proposal of Ji to calculate quasi-distributions~\cite{Ji:2013dva,Ji:2014gla}, a lot of theoretical and numerical effort has been put to understand and extend this and related approaches, see e.g.~Refs.~\cite{Cichy:2018mum,Ji:2020ect,Constantinou:2020pek,Cichy:2021lih,Cichy:2021ewm} for recent reviews.
Most of the reported progress concerns parton distribution functions (PDFs), but more recently, lattice extraction of GPDs also received a lot of attention.
Results were reported for the matching~\cite{Ji:2015qla, Xiong:2015nua, Liu:2019urm, Radyushkin:2019owq, Ma:2022ggj}, within phenomenological models~\cite{Bhattacharya:2018zxi, Bhattacharya:2019cme, Ma:2019agv, Luo:2020yqj, Shastry:2022obb} and, particularly, first pioneering computations were obtained for the pion GPDs~\cite{Chen:2019lcm} and the nucleon GPDs -- unpolarized and helicity~\cite{Alexandrou:2020zbe}, transversity~\cite{Alexandrou:2021bbo} and even twist-3 ones~\cite{Bhattacharya:2021oyr}.

These proceedings is a continuation of Ref.~\cite{Shohini_Lattice}, which was presented at the same conference, and addresses a new approach to access GPDs from any kinematic frame. The approach was proposed in Ref.~\cite{Bhattacharya:2022aob}, to which we refer the reader for the details of the methodology and its implementation. According to the new approach, the proton matrix elements for the vector operator are parametrized in eight linearly-independent Dirac structures, each accompanied by a Lorentz-invariant amplitude $A_i$, that is 
\vspace*{-0.1cm}
\begin{align}
\label{eq:parametrization_general}
F^{\mu} (z,P,\Delta) & = \bar{u}(p_f,\lambda') \bigg [ \dfrac{P^{\mu}}{m} A_1 + m z^{\mu} A_2 + \dfrac{\Delta^{\mu}}{m} A_3 + i m \sigma^{\mu z} A_4 + \dfrac{i\sigma^{\mu \Delta}}{m} A_5 \nonumber \\
& \hspace{1.8cm} + \dfrac{P^{\mu} i\sigma^{z \Delta}}{m} A_6 + m z^{\mu} i\sigma^{z \Delta} A_7 + \dfrac{\Delta^{\mu} i\sigma^{z \Delta}}{m} A_8  \bigg ] u(p_i, \lambda) \, ,\,\, \mu=0,1,2,3.
\end{align}
\vskip -0.2cm
Here, we perform a proof-of-concept analysis of the unpolarized GPDs, as extracted from the symmetric:
$\vec{p}^{\,s}_f{=}\vec{P} {+} {\vec{\Delta}}/{2}$, 
$\,\,\vec{p}^{\,s}_i{=}\vec{P} {-} {\vec{\Delta}}/{2}$,
and asymmetric:
$\vec{p}^{\,a}_f{=}\vec{P}$,
$\,\,\vec{p}^{\,a}_i{=}\vec{P} {-} \vec{\Delta}$ frames, where $\vec{\Delta}{=} \left(\Delta_1,\Delta_2,0\right)$ (zero skewness, $\xi{=}0$).
We compare the extracted $A_i$ between frames and assess numerically their Lorentz invariance. A numerical confirmation is a highly non-trivial check of the numerical calculations. We emphasize that the asymmetric frame is computationally more efficient, as one can obtain more than one value of $\vec{\Delta}$ within the same computational cost. 
We calculate the matrix elements of the non-local vector operator containing a Wilson line. The proton states are momentum-boosted with nonzero momentum transfer between the initial ($|N(P_i)\rangle$) and final ($|N(P_f)\rangle$) state, $\langle N(p_f)|\bar\psi\left(z\right) \gamma_j {\cal W}(0,z)\psi\left(0\right)|N(p_i)\rangle\,.$
These are constructed by an optimized ratio between the 3pt- and 2pt-functions. The ground state of the matrix element is extracted from a single-state fit with respect to the operator insertion time and is denoted by ${\Pi_\mu}$. The expressions for the parity-projected matrix elements are very lengthy and are given in Ref.~\cite{Bhattacharya:2022aob}. These are the basis of matrix elements to disentangle the eight Lorentz invariant amplitudes. The amplitudes are directly related to the quasi-GPDs of $H$ and $E$, which are not uniquely defined. Here, we use the $\gamma^0$ definition in the symmetric and asymmetric frames, as well as a Lorentz-invariant definition. With the $A_i$ being frame-independent, one can relate the quasi-GPDs to the matrix elements of either frame; this is a powerful relation, as a calculation in the asymmetric frame requires fewer computational resources. With this in mind one can use the ``standard'' definition for the quasi-GPDs,
\begin{align}
F^{\mu} (z,P,\Delta) & = \bar{u}(p_f,\lambda') \bigg [ \gamma^{0} {\cal{H}}_0 (z,P,\Delta) + \frac{i\sigma^{0\mu}\Delta_{\mu}}{2m} {\cal{E}}_0 (z,P,\Delta) \bigg] u(p_i, \lambda) \,,
\label{e:historic}
\end{align}
\noindent in either frame, leading to the following relations for zero skewness:
\begin{eqnarray}
\label{eq:FHs}
{\cal H}_0^s(A_i^s;z) \hspace*{-0.15cm} & = & \hspace*{-0.15cm}
A_1 + \frac{z (\Delta_1^2 + \Delta^2_2)}{2 P_3}  {{A_6}} \,,\\[3ex]
\label{eq:FHa}
{\cal E}_0^s(A_i^s;z) \hspace*{-0.15cm} & = & \hspace*{-0.15cm}
- A_1 - \frac{m^2 z }{P_3}  {{A_4}} + 2 A_5 - \frac{z \left( 4 E^2 + \Delta_1^2 + \Delta^2_2\right)}{2P_3}   {{A_6}}   \,,\\[3ex]
\label{eq:FHa}
{\cal H}_0^a(A_i^a;z) \hspace*{-0.15cm} & = & \hspace*{-0.15cm}
A_1 + \frac{\Delta_0}{P_0} {{A_3}} + \frac{m^2 z \Delta_0}{2P_0P_3}  {{A_4}}+ \frac{z (\Delta_0^2 + \Delta_1^2+ \Delta_2^2)}{2 P_3}  {{A_6}} + \frac{z (\Delta_0^3 + \Delta_0 (\Delta_1^2+ \Delta_2^2))}{2P_0 P_3}   {{A_8}} \,,\\[3ex]
\label{eq:FEa}
{\cal E}_0^a(A_i^a;z)  \hspace*{-0.15cm} & = & \hspace*{-0.15cm}
- A_1 - \frac{\Delta_0}{P_0}  {{A_3}} - \frac{m^2 z (\Delta_0+2P_0)}{2P_0P_3}  {{A_4}} + 2 A_5 - \frac{z \left(\Delta_0^2 + 2 P_0 \Delta_0 + 4 P_0^2 + \Delta_1^2+ \Delta_2^2 \right)}{2P_3}   {{A_6}} \nonumber\\[2ex] 
&&  - \frac{z \Delta_0 \left(\Delta_0^2 + 2 \Delta_0 P_0 + 4 P_0^2 + \Delta_1^2+ \Delta_2^2\right)}{2P_0 P_3}   {{A_8}}\,.
\end{eqnarray}
We emphasize again that one may use either definition for the quasi-GPDs employing the Lorentz-invariant amplitudes from any frame of choice. For instance, one can extract ${\cal H}_0^s$ and  ${\cal E}_0^s$ from the asymmetric frame $A_i^a$, as the symmetric frame is computationally costly and not optimal for lattice QCD calculations. We note that (${\cal H}_0^s$, ${\cal E}_0^s$) differ from (${\cal H}_0^a$, ${\cal E}_0^a$) at finite momentum $P_3$ due to their Lorentz non-invariant definition. In the infinite momentum limit, both approach the correct light-cone GPDs. Another aspect of this work is a new Lorentz-invariant definition of quasi-GPDs, ${\cal H}$ and ${\cal E}$, that may have faster convergence to light-cone GPDs. Being Lorentz invariant, these are equivalent in both frames, and at zero skewness, one obtains
\begin{eqnarray}
\label{eq:FH_impr}
{\cal H}(A_i^{s/a};z)   =  A_1 \,,\quad
{\cal E}(A_i^{s/a};z)   =  - A_1 + 2 A_5 + 2 z P_3 A_6 \,.
\end{eqnarray}
We note that the definition of ${\cal H}$ and ${\cal E}$ can be interpreted as the construction of a new operator that is a combination of $\gamma_0$, $\gamma_1$ and $\gamma_2$.
We emphasize that it is important to provide a comparison of the ${\cal H}$ and ${\cal E}$ GPDs in the two frames at the same value of $t$, which would require $\Delta^{a}_\perp \neq \Delta^{s}_\perp$. Such a relation is $P_3$-dependent, but for the values of $P_3$ employed in this work $t$ are numerically similar (Table~\ref{tab:stat}).

The calculation is performed on an $N_f=2+1+1$ gauge ensemble of  twisted-mass fermions~\cite{Alexandrou:2018egz} with volume $32^3 \times 64$ and lattice spacing $a=0.093$ fm, in which 
the pion mass is 260 MeV. We use a source-sink separation of $t_s=10 a = 0.93$ fm to keep the statistical noise under control. We employ momentum smearing~\cite{Bali:2016lva} to improve the overlap with the proton ground state and suppress gauge noise, as demonstrated in Ref.~\cite{Alexandrou:2016jqi} for non-local operators.
In Table~\ref{tab:stat}, we give the statistics for the symmetric and asymmetric frames. 
{\small{
\begin{table}[h!]
\begin{center}
\begin{tabular}{lcccc|cccc}
\hline
frame & $P_3$ [GeV] & $\quad \mathbf{\Delta}$ $[\frac{2\pi}{L}]\quad$ & $-t$ [GeV$^2$] & $\quad \xi \quad $ & $N_{\rm ME}$ & $N_{\rm confs}$ & $N_{\rm src}$ & $N_{\rm tot}$\\
\hline
symm      & $\pm$1.25 &($\pm$2,0,0), (0,$\pm$2,0)  &0.69   &0   &8   &249 &8  &15936 \\
asymm  & $\pm$1.25 &($\pm$2,0,0), (0,$\pm$2,0)  &0.64   &0   &8   &269 &8  &17216\\
\hline
\end{tabular}
\vspace*{-.15cm}
\caption{Calculation parameters and statistics. $N_{\rm ME}$, $N_{\rm confs}$, $N_{\rm src}$ and $N_{\rm total}$ is the number of matrix elements, configurations, source positions per configuration and total statistics, respectively.}
\label{tab:stat}
\end{center}
\end{table}
}}

\vspace*{-0.5cm}
\section{Lattice results}

\subsection{Matrix elements and quasi-GPDs}

\begin{figure}[h!]
    \centering
    \includegraphics[scale=0.205]{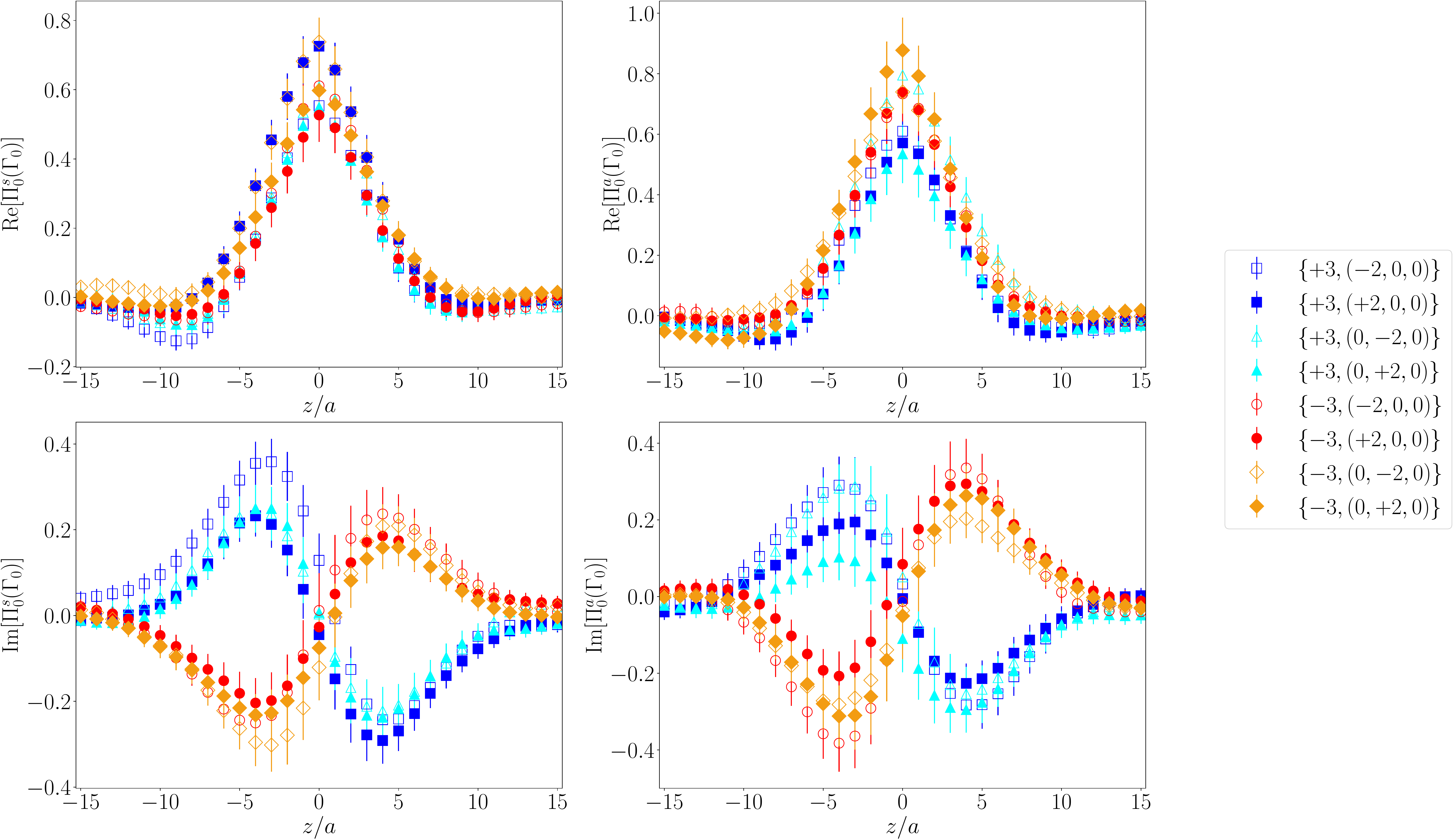}
    \vspace*{-0.2cm}
    \caption{\small{Bare matrix element $\Pi_0(\Gamma_0)$ in the symmetric frame (left) and in the asymmetric frame (right), for $|P_3|=1.25$ GeV and $t=-0.69$ GeV$^2$ ($t=-0.64$ GeV$^2$) for the symmetric (asymmetric) frame. The top (bottom) panel corresponds to the real (imaginary) part. The notation in the legend is $\frac{L}{2\pi}\{P_3,\vec{\Delta}\}$. }}
    \label{fig:Pi0G0}
\end{figure}

In this section, we provide selected results in the two frames. An extended discussion can be found in Ref.~\cite{Bhattacharya:2022aob}.
In Fig.~\ref{fig:Pi0G0}, we show the bare $\Pi_0(\Gamma_0)$ (unpolarized parity projector) in both frames for all combinations of $\pm P_3$ and $\pm \vec{\Delta}$ given in Table~\ref{tab:stat}. 
The matrix elements in the symmetric frame have definite symmetries with respect to $P_3 \to -P_3$, $z \to -z$, and $\vec{\Delta} \to -\vec{\Delta}$, unlike the case of the asymmetric frame. Therefore, any combination of $P_3 \to -P_3$, $z \to -z$, and $\vec{\Delta} \to -\vec{\Delta}$ is handled at the level of $A_i$ that have definite symmetries (see Appendix B of Ref.~\cite{Bhattacharya:2022aob}). 
Quantitative comparison of $\Pi^s_\mu$ and $\Pi^a_\mu$ is not meaningful, as the matrix elements are not equivalent. For instance, $\Pi^s_0(\Gamma_0)$ contains information on $A_1$, $A_5$, and $A_6$, while $\Pi^a_0(\Gamma_0)$ decomposes to $A_1$, $A_3$, $A_4$, $A_5$, $A_6$, and $A_8$. 
We find that the lack of symmetries in $\Pi^a_0(\Gamma_0)$ is a small effect numerically. The same holds for other matrix elements in the asymmetric frame, such as $\Pi^a_2(\Gamma_3)$, but not $\Pi^a_1(\Gamma_2)$.

\begin{figure}[h!]
    \centering
    \includegraphics[scale=0.21]{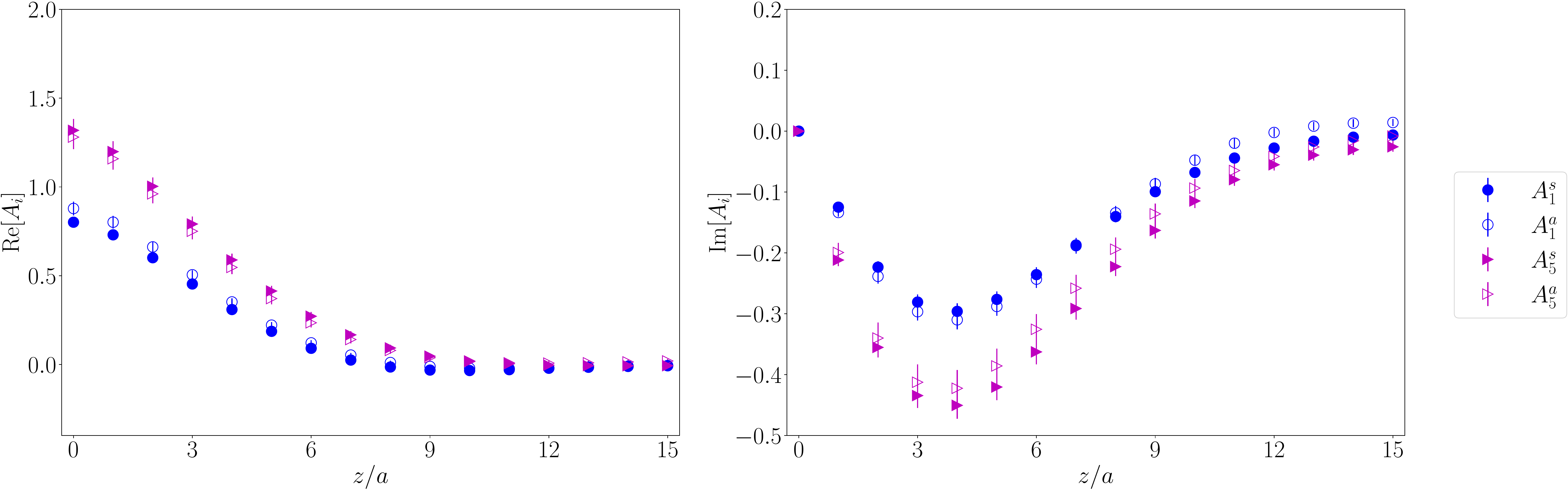}
    \vspace*{-0.3cm}
    \caption{Comparison of bare $A_1$ and $A_5$ in the symmetric (filled symbols) and asymmetric (open symbols) frame. The real (imaginary) part of each quantity is shown in the left (right) column.}
    \label{fig:A156}
\end{figure}

We decompose the Lorentz-invariant amplitudes using the matrix elements of $\gamma^\mu$ ($\mu=0,1,2,3$) with the unpolarized and polarized parity projectors. The fact that the $A_i$ are frame-invariant makes them interesting to study closely before extracting the quasi-GPDs. A numerical test of their frame independence is a consistency check of the lattice estimates for $A_i$. The level of agreement in the two frames provides an estimate of systematic uncertainties. Here, we present the $A_i$ after we apply the appropriate symmetrization with respect to $\pm P_3 z$. This allows us to have better statistical accuracy by about $1/\sqrt{8}$. In Fig.~\ref{fig:A156}, we present the bare $A_1$ and $A_5$; the remaining amplitudes are found to be very small or negligible, which is due to the small magnitude of some matrix elements, such as $\Pi_1(\Gamma_2)$. We find that $A_5$ has the largest magnitude, followed by $A_1$. Overall, we find very good agreement between the two frames for each $A_i$. We remind the reader that there is 7\% difference in the momentum transfer between the two frames (${-}t^s{=}0.69$ GeV$^2$, ${-}t^a{=}-0.64$ GeV$^2$), which may be responsible for the small differences seen in Fig.~\ref{fig:A156}. Such a difference between $t^s$ and $t^a$ is, in general, not an obstacle in our approach, as a Lorentz boost transformation can relate the momentum transfer between the two frames, without ambiguity in the extracted light-cone GPDs.

\begin{figure}[h!]
    \centering
    \includegraphics[scale=0.23]{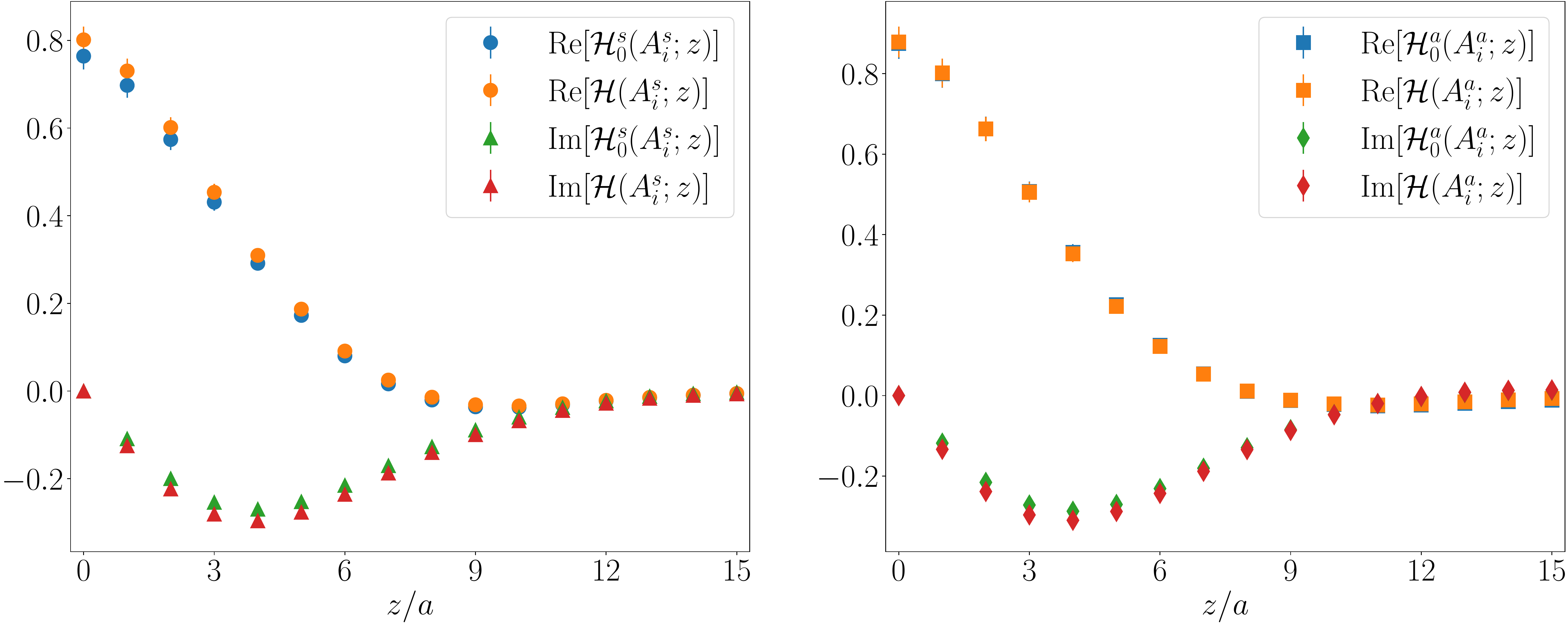}\\
        \,\,\,\,\includegraphics[scale=0.23]{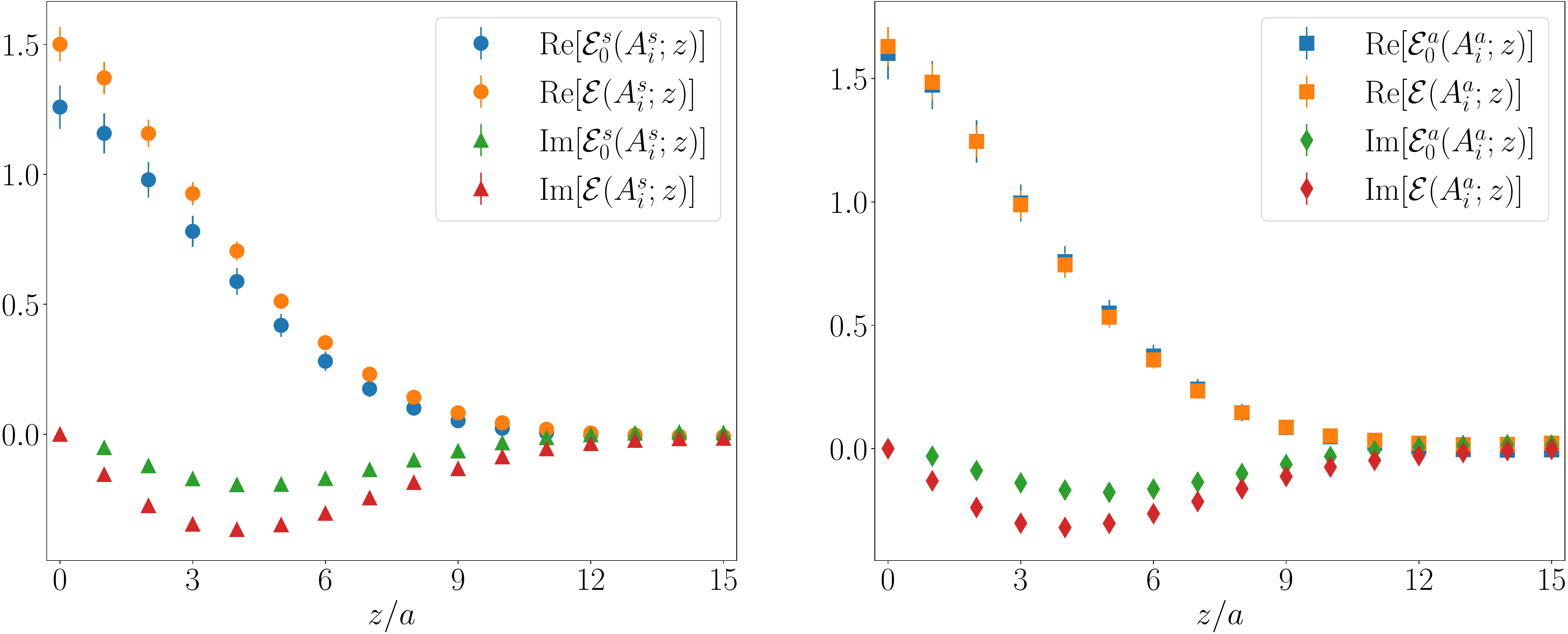}
     \vspace*{-.3cm}
    \caption{Top: Comparison of bare ${\cal H}_0$ and ${\cal H}$ at $|P_3|=1.25$ GeV in the symmetric (left, $t=-0.69$ GeV$^2$) and asymmetric (right, $t=-0.64$ GeV$^2$) frame. Bottom: Similar to top plots but for ${\cal E}_0$ and ${\cal E}$.}
    \label{fig:FH_a}
\end{figure}

Using the $A_i$, we extract the quasi-GPDs from the three definitions mentioned previously. The results for the $\gamma^0$ definition for $P_3{=}1.25$ GeV and $t^s{=-}0.69$ GeV$^2$, $t^a{=-}0.64$ GeV$^2$ are shown in Fig.~\ref{fig:FH_a}. 
In particular, we compare the standard, ${\cal H}_0$ and ${\cal E}_0$,  and Lorentz invariant definitions, ${\cal H}$ and ${\cal E}$, as calculated in each frame. We remind the reader that defining ${\cal H}_0$ and ${\cal E}_0$ through $\gamma_0$ is frame-dependent and, thus, ${\cal H}_0^s$ and ${\cal H}_0^a$ have a different functional form; similarly for ${\cal E}_0^s$, ${\cal E}_0^a$. 
We find that for this kinematic setup, ${\cal H}_0$ is fully compatible with ${\cal H}$ in both frames. An excellent agreement is found between ${\rm Re}[{\cal E}]$ and ${\rm Re}[{\cal E}_0]$ in the asymmetric frame, while in the symmetric frame, there are some differences. Differences are also observed between ${\rm Im}[{\cal E}]$ and ${\rm Im}[{\cal E}_0]$ for both frames. 
It is interesting to observe that the statistical errors are considerably smaller in ${\cal E}$ as compared to ${\cal E}_0$.
Tracing this behavior in the raw data, the definition of ${\cal E}$ involves additional matrix elements that subtract the noise present in $\Pi_0(\Gamma_{1/2})$. The statistical accuracy in the case of $H$ is similar across definitions and frames. Finally, we find that the Lorentz-invariant quasi-GPDs from the two frames are in agreement, as expected theoretically.

An important component of the lattice calculation is renormalization, for which we use an RI-type prescription. In particular, we implement the standard RI prescription defined on a single renormalization scale, $(a  \mu_0)^2\approx1.95$, for compatibility with the matching formalism of Refs.~\cite{Liu:2019urm,LatticeParton:2018gjr}. 
As discussed in Ref.~\cite{Bhattacharya:2022aob}, the appropriate renormalization for $H$ and $E$ is that of $\gamma_0$, which is valid for both Lorentz-invariant and non-invariant quasi-GPDs. 
Details on the calculation of the renormalization functions used in this work can be found in Ref.~\cite{Alexandrou:2019lfo}. 
It is well-known that there are challenges related to renormalization, that is, as $z$ increases, the RI prescriptions become less reliable. 
In practice, the value of the renormalization functions increases exponentially due to the linear divergence leading to renormalized functions that do not decay to zero. Alternative renormalization prescriptions, such as the hybrid scheme~\cite{Ji:2020brr}, and reduction of lattice artifacts in the RI estimates~\cite{Constantinou:2022aij} may help alleviate the problem.

\subsection{Light-cone GPDs}

The light-cone GPDs are extracted by reconstructing the $x$-dependence of the quasi-GPDs and then applying the appropriate matching equations. The reconstruction is challenging due to the limited number of lattice data leading to the so-called inverse problem~\cite{Karpie:2019eiq}, which mainly affects the small-$x$ region. The moderate-to-large-$x$ region is not sensitive to this inverse problem, thus allowing us to make reliable predictions.
We use the Backus-Gilbert (BG) reconstruction method~\cite{BackusGilbert}, which uses a model-independent criterion to select the light-cone GPDs from among the infinite set of possible solutions to the inverse problem: the variance of the solution with respect to the statistical variation of the input data should be minimal. In this work, we vary the number of data that enter the reconstruction, that is, $z_{\rm max}=7a,\,9a,\,11a,\,13a$. We find little sensitivity in the value of  $z_{\rm max}$ in the reconstructed data, and use as final estimates the ones obtained from $z_{\rm max}=9a$. For the matching, we use the one-loop expression of Ref.~\cite{Liu:2019urm} to extract the light-cone GPDs in the $\overline{\rm MS}$ scheme at 2 GeV. 
At zero skewness, the matching coefficient is the same as in the quasi-PDF case~\cite{Liu:2019urm}.

\begin{figure}[h!]
    \centering
    \includegraphics[scale=0.22]{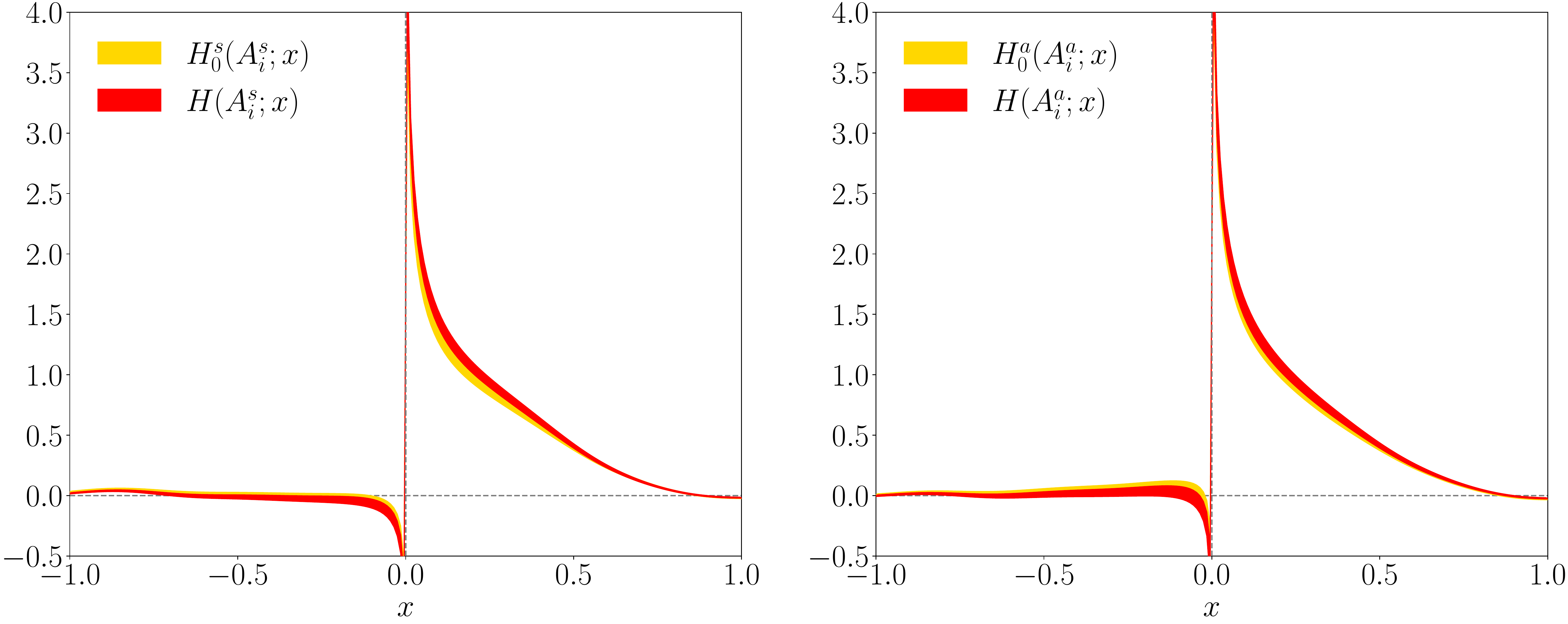}\\
        \includegraphics[scale=0.22]{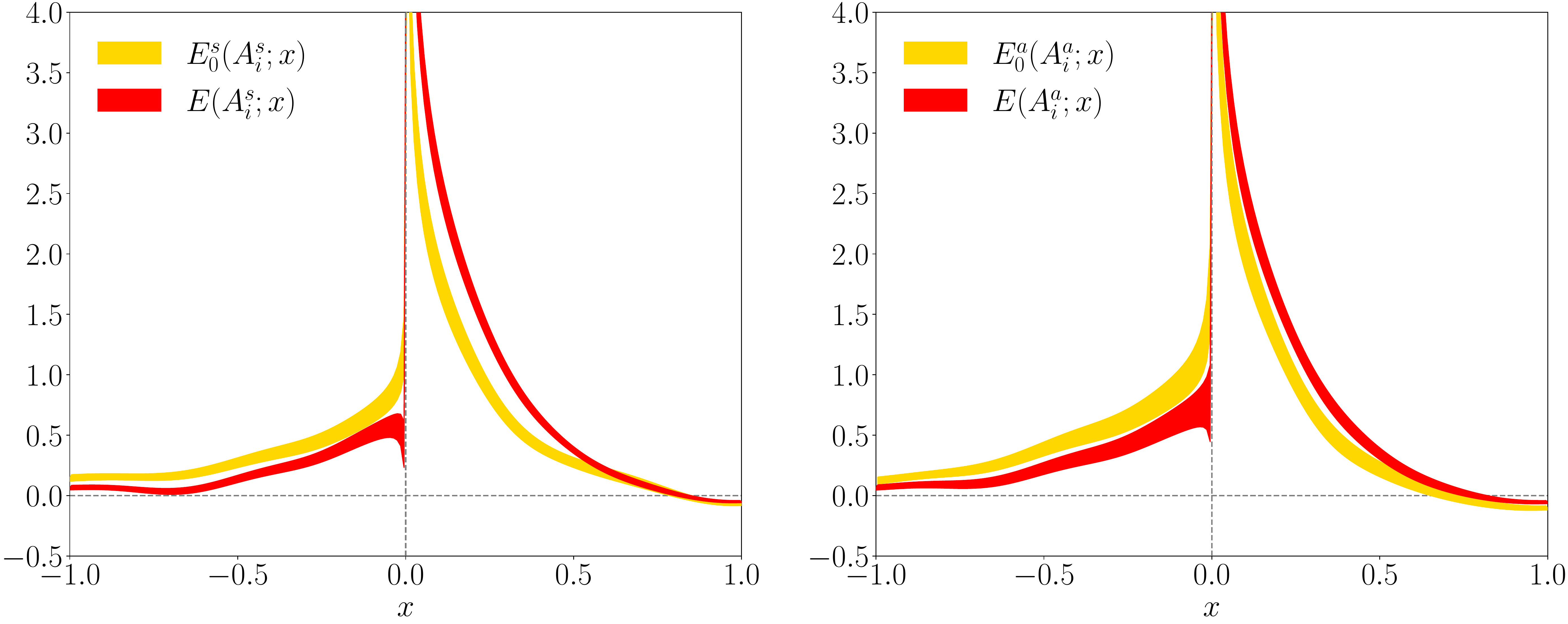}   
    \vspace*{-0.3cm}
    \caption{Comparison of light-cone $H_0$ and $H$ (top) and $E_0$ and $E$ (bottom) in the symmetric (left, $t=-0.69$ GeV$^2$) and asymmetric (right, $t=-0.64$ GeV$^2$) frame. Results are presented in the $\overline{\rm MS}$ scheme at 2 GeV.}
    \label{fig:H_calH}
\end{figure}
\begin{figure}[h!]
    \centering
    \includegraphics[scale=0.22]{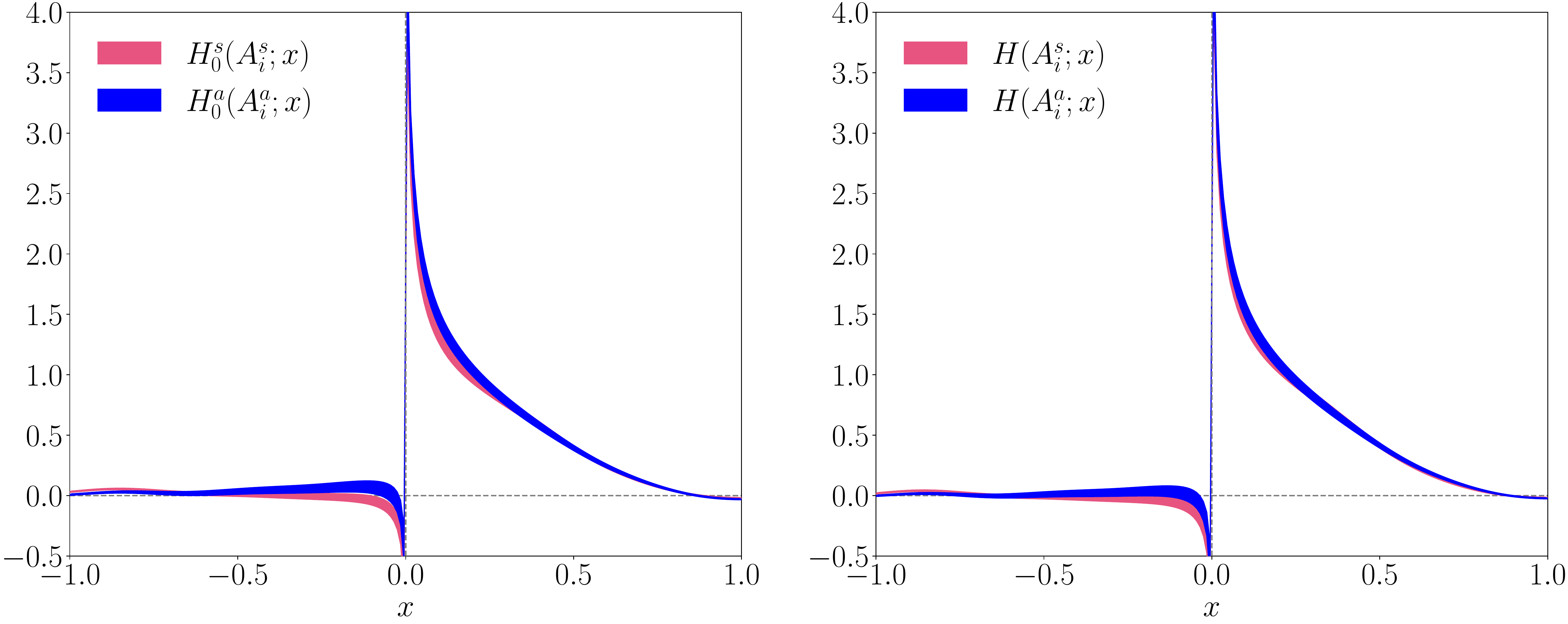}\\
        \includegraphics[scale=0.22]{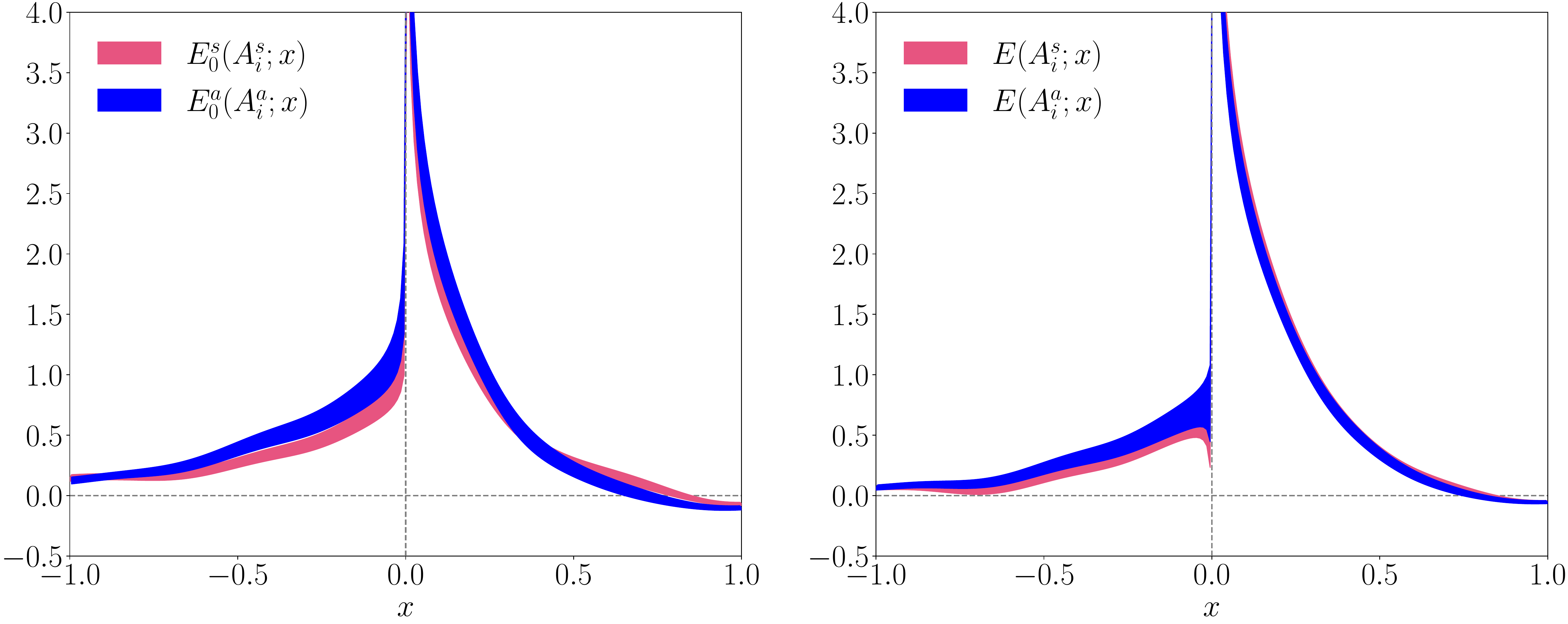}
    \vspace*{-0.3cm}
    \caption{Left: Comparison of light-cone $H_0$ (top) and $E_0$ (bottom) in the symmetric ($t=-0.69$ GeV$^2$) and asymmetric ($t=-0.64$ GeV$^2$) frame. Right: Comparison of light-cone $H$ (top) and $E$ (bottom) in the symmetric and asymmetric frame. Results are presented in the $\overline{\rm MS}$ scheme at 2 GeV.}
    \label{fig:Hs_Ha}
\end{figure}

In Fig.~\ref{fig:H_calH} - \ref{fig:Hs_Ha}, we compare the light-cone GPDs, as extracted from different definitions and frames.
In particular, Fig.~\ref{fig:H_calH} demonstrates that the Lorentz invariant and non-invariant definitions for the $H$-GPD are the same for both the symmetric and the asymmetric frames. 
Unlike the case of the $H$-GPD, the two definitions of $E$-GPD are of similar magnitude and shape but are not in agreement for most of the $x$ region. Note that they are expected to be different by construction. Interestingly, the numerical difference between $E_0^s$ and $E$ is more prominent in the symmetric frame. 
In addition to comparing the results from different definitions within the same frame, it is interesting to investigate whether the two frames for a given definition show any agreement. 
An agreement between different frames is expected theoretically only for the Lorentz-invariant definitions, $H$ and $E$. 
Indeed, Fig.~\ref{fig:Hs_Ha} confirms perfect agreement in both the $H$ and $E$ in the two frames. Furthermore, such an agreement is also observed in the Lorentz non-invariant definitions $H$ and $E$.

\section{Summary}
\label{sec:summary}

Calculations of $x$-dependent GPDs within lattice QCD are usually assumed in the symmetric kinematic frame, which has the disadvantage of high computational cost to obtain them in a wide range of $t$ and $\xi$. The main motivation of the approach described here is to calculate the GPDs in a computationally efficient manner. In particular, we implement a new method to parameterize the off-forward matrix elements relevant to GPDs in terms of Lorentz-invariant amplitudes~\cite{Bhattacharya:2022aob}. The method is applicable for all operators, but here we focus on the unpolarized GPDs (vector operator). The parameterization is such that the frame dependence of the matrix elements is absorbed in the kinematic factors of Lorentz-invariant amplitudes, $A_i$.
We calculate in lattice QCD the $A_i$ in two frames, and we find them to be frame-independent.

We reiterate that the use of the new parametrization in any frame significantly reduces the computational cost. In the fixed-sink sequential inversion approach, the asymmetric setup presented here is at least four times less costly than the symmetric frame calculation. For instance, one can quadruple the number of measurements by adding all permutations of $\vec{\Delta}$ contributing to the same $t$. Also, several vectors $\vec{\Delta}$ may be obtained for a given $\vec{p}_f$ with an overhead of only the contraction cost. A preliminary analysis for various values of $t$ obtained at once shows a good signal for several values.

The Lorentz-invariant amplitudes are directly related to the quasi-GPDs of $H$ and $E$. The latter are not uniquely defined, and we investigated three definitions:
(a) symmetric frame via the $\gamma_0$ operator (${\cal H}_0^s,\,{\cal E}_0^s$);
(b) asymmetric frame via the $\gamma_0$ operator (${\cal H}_0^a,\,{\cal E}_0^a$);
(c) Lorentz-invariant ($\cal H,\,E$).
These definitions are not equivalent, as they differ by power corrections.
${\cal H}_0^s,\,{\cal E}_0^s$ are of particular interest, as they have been used in previous lattice QCD calculations in the symmetric frame. 
One may extract the quasi-GPDs in the symmetric frame in a computationally less-costly way by using the Lorentz-invariant amplitudes $A_i$ obtained from the asymmetric frame. 
In such a case, the quasi-GPDs are defined at the value of $t$ corresponding to the asymmetric frame setup. Exploration of the Lorentz-invariant definition is also interesting. Here, we confirm numerically the frame independence of $\cal H$ and $\cal E$.
In closing, the proposed parametrization and the introduction of the Lorentz-invariant amplitudes is a powerful theoretical tool and has a broad range of interesting applicability that we will explore in the future.


\vspace{0.75cm}

\centerline{\textbf{Acknowledgements} }

\vspace{0.25cm}
This material is based upon work supported by the U.S. Department of Energy, Office of Science, Office of Nuclear Physics through Contract No.~DE-SC0012704, No.~DE-AC02-06CH11357 and within the framework of Scientific Discovery through Advance Computing (SciDAC) award Fundamental Nuclear Physics at the Exascale and Beyond (S.~B. and S.~M.).
K.~C.\ is supported by the National Science Centre (Poland) grants SONATA BIS no.\ 2016/22/E/ST2/00013 and OPUS no.\ 2021/43/B/ST2/00497. M.~C., J. D. and A.~S. acknowledge financial support by the U.S. Department of Energy, Office of Nuclear Physics, Early Career Award under Grant No.\ DE-SC0020405.
 J. D. also received support by the U.S. Department of Energy, Office of Science, Office of Nuclear Physics, within the framework of the TMD Topical Collaboration. 
The work of A.~M. has been supported by the National Science Foundation under grant number PHY-2110472, and also by the U.S. Department of Energy, Office of Science, Office of Nuclear Physics, within the framework of the TMD Topical Collaboration. 
F.~S.\ was funded by by the NSFC and the Deutsche Forschungsgemeinschaft (DFG, German Research Foundation) through the funds provided to the Sino-German Collaborative Research Center TRR110 “Symmetries and the Emergence of Structure in QCD” (NSFC Grant No. 12070131001, DFG Project-ID 196253076 - TRR 110). 
 YZ was partially supported by an LDRD initiative at Argonne National Laboratory under Project~No.~2020-0020.
Computations for this work were carried out in part on facilities of the USQCD Collaboration, which are funded by the Office of Science of the U.S. Department of Energy. 
This research used resources of the Oak Ridge Leadership Computing Facility, which is a
DOE Office of Science User Facility supported under Contract DE-AC05-00OR22725.
This research was supported in part by PLGrid Infrastructure (Prometheus supercomputer at AGH Cyfronet in Cracow).
Computations were also partially performed at the Poznan Supercomputing and Networking Center (Eagle supercomputer), the Interdisciplinary Centre for Mathematical and Computational Modelling of the Warsaw University (Okeanos supercomputer), and at the Academic Computer Centre in Gda\'nsk (Tryton supercomputer). The gauge configurations have been generated by the Extended Twisted Mass Collaboration on the KNL (A2) Partition of Marconi at CINECA, through the Prace project Pra13\_3304 ``SIMPHYS".
Inversions were performed using the DD-$\alpha$AMG solver~\cite{Frommer:2013fsa} with twisted mass support~\cite{Alexandrou:2016izb}. 


\bibliographystyle{h-physrev}

\bibliography{references.bib}

\end{document}